\documentclass[prb,twocolumn,showpacs,floats]{revtex4}
\usepackage{amsmath,color,graphicx,amsbsy}

\newcommand{\curl}{{\boldsymbol{\nabla}\times}}
\newcommand{\bold}[1]{{\boldsymbol{#1}}}

\begin{document}

\title{Considerations on the symmetry of loop order in cuprates and some effects accompanying it.}
\author{A.~Shekhter$^1$ and C.~M.~Varma$^2$}
\affiliation{
$^1$National High Magnetic Field Lab, Tallahassee, Florida 32310\\
$^2$Department of Physics and Astronomy, University of California, Riverside, California 92521}
\begin{abstract} 
	The loop-current state discovered in the pseudogap phase of cuprates breaks time reversal symmetry and lowers the point group symmetry of the crystal. The order parameter and the magnetic structure within each unit cell which is associated with it can be described by a toroidal moment parallel to the copper-oxide planes. We discuss lattice point group symmetry of the magnetic structure. As an application, we discuss a few effects that necessarily accompany order parameter in the pseudogap phase. The magnitude estimated for these specific effects makes them hard to observe because they rely on the small magnetic fields associated with the order parameter. Effects, associated with the electronic energies are much larger. Some of them have already been discussed.
\end{abstract}
\date{\today}

\maketitle

\section{Introduction.}

Polarized neutron scattering experiments, dichroic angle resolved photoemission experiments and magnetization measurements \cite{NeutronExp,greven,Kaminski,Monod2008,simon-varma} reveal the predicted line of  phase transition \cite{varma} in cuprates at pseudogap temperature $T^*(x)$. At the present time four classes of cuprates have been studied, all of which show consistency with the same form of order.  The purpose of this paper is to present a symmetry analysis of the order parameter, which is necessary for further work. The neutron spin-flip intensity observed in the experiment in the pseudogap phase is consistent with the magnetic structure that transforms under the two-dimensional irreducible representation $E_u$ of the point group of the copper-oxide lattice, $D_{4h}$. The order parameter  ${\bf L}$ that can be used to characterize the magnetic structure of such symmetry is a polar time-reversal-odd vector parallel to the copper-oxide planes. 
Such a vector has been termed a "toroidal" or an "anapole" moment in the literature.
The relation of such an order parameter to the magnetic structure and the loop-currents in the unit cell of cuprates is specified. 

Several experiments other than those which measure the order parameter directly \cite{NeutronExp,greven,Kaminski,Monod2008,simon-varma} have already been proposed which lead to unusual effects due to the coupling of the order parameter to external probes. These include several forms of dichroism in x-ray scattering \cite{dimatteo} and second harmonic generation in optical experiments \cite{simon}. As an application of the symmetry analysis presented for the first time in this paper, we discuss few other physical phenomena that necessarily accompany the order parameter $\bold{L}$. These are crystal lattice distortions which are second order in the order parameter, crystal distortion linear in an applied uniform field and linear in the order parameter, and magneto-electric phenomena in transport properties. These are effects related to the small energies associated with the magnetic field due to the order parameter. Much larger effects are associated with the electronic energies associated with the order parameter. This distinction is similar to the difference between the exchange energies and energies of magnetic fields associated with spin-moment order. For example, the spin-splitting in ferromagnetic iron (ordered moment of $\approx 2.5 \mu_B$/unit-cell) is about 1 eV, while the magnetostrictive distortion due to the magnetic fields is  a lattice distortion only about 1 part in $10^3$.

\section{Loop order parameter: Toroidal moment.}

The polarized neutron scattering experiments\cite{NeutronExp,greven} reveal an elastic spin-flip intensity on top of the subset of Bragg peaks in pseudogap phase in cuprates. No new Bragg peaks appear; the magnetic structure responsible for spin-flip intensity does not break translational invariance. However, it follows from the analysis of the neutron data that the symmetry of the observed magnetic structure is lower than the point group symmetry of the lattice. The copper-oxide lattice has tetragonal symmetry 4/mmm ($D_{ 4h}$) (the effects of the small orthorhombic distortion due to ordering  in Cu-O chains in some cuprates will be mentioned). In the pseudogap phase the symmetry is lowered to  $\underline{m}mm$ or $D_{2h}(C_{2v})$ \cite{simon-varma,dimatteo}. Such lowering of the symmetry follows if the magnetic structure transforms under irreducible representation  $E_u$ of the point group of the lattice. For an order parameter in the pseudogap phase one can choose a polar in-plane (parallel to copper-oxide plane) vector  $\boldsymbol{L}=(L_{y}, L_{x})$ which is restricted to four crystalline directions because it  transforms the same way, $E_u$,  under the operations of point group of the lattice. Order parameter $\bold{L}$ is time-reversal-odd because the magnetic structure which it represents is odd under time reversal. In the literature polar time-reversal-odd vector is called a toroidal moment as it is a symmetry of a magnetic field in a solenoid bent into a torus \cite{PhysRep}; in the particle physics object of the same symmetry is known as anapole moment \cite{zeldovich-1958}.

A natural way to relate the order parameter $\boldsymbol{L}$ to the pattern of the magnetization $\bold{M}(\bold{r})$ in the pseudogap phase is as follows. Since the observed magnetic structure retains lattice translation symmetries it is enough to consider magnetization  $\bold{M}(\bold{r})$ within single unit cell. In general, the function $\bold{M}(\bold{r})$ can be decomposed into spatial harmonics of point group of the lattice  ($D_{4h}$);  $\bold{M}(\bold{r})=\sum_{\lambda\alpha}c_{\lambda\alpha}\bold{M}_{\lambda\alpha}(\bold{r})$ .  Here harmonic $\bold{M}_{\lambda\alpha}(\bold{r})$ transforms under the irreducible representation $\lambda$ of the group $D_{4h}$ and $\alpha$ is internal index of the representation $\lambda$ if it is not one-dimensional; $c_{\lambda\alpha}$ is a set of coefficients. The mathematical representation of the fact that the magnetic structure $\bold{M}(\bold{r})$ belongs to the irreducible representation $E_u$ is that it can be written as $\bold{M}(\bold{r})=c_1\bold{M}_1(\bold{r})+c_2\bold{M}_2(\bold{r})$ where $\bold{M}_{1,2}(\bold{r})$ are two orthogonal harmonics that transform under two-dimensional irreducible representation $E_u$. It can be seen that under all group operations a pair of vectors  $\bold{L}_{\alpha=1,2}$ defined by 
\begin{align}\label{eq:toroidal-alpha} 
\bold{L}_{\alpha} = \int\limits_{\text{unit cell}} d\bold{r} \bold{M}_{\alpha}(\bold{r})\times\bold{r}
\end{align}
also transforms under $E_u$ representation. The two vectors $\bold{L}_{\alpha}$ are parallel to the copper-oxide plane and are orthogonal to each other; we can use them as a basis to define a toroidal moment $\bold{L}=c_1\bold{L}_1+c_2\bold{L}_2$ which is associated with the magnetic structure  $\bold{M}(\bold{r})=c_1\bold{M}_1(\bold{r})+c_2\bold{M}_2(\bold{r})$ in the unit cell. 
Conversely, if the magnetization  $\bold{M}(\bold{r})$ does not have a component that transforms under  $E_u$, the integral in Eq.~(\ref{eq:toroidal-alpha}) vanishes.   We also note that the out-of plane component of the integral in Eq.~(\ref{eq:toroidal-alpha}), if non-zero, transforms under $A_{2u}$ representation; it vanishes if $\bold{M}_{1,2}(\bold{r})$ transforms under  $E_u$ representation. We conclude that the vector $\bold{L}$ defined by 
\begin{align}\label{eq:toroidal} 
\bold{L} = \int\limits_{\text{unit cell}} d\bold{r} \bold{M}(\bold{r})\times\bold{r}
\end{align}
can be used as an order parameter to characterize the magnetic structure in the pseudogap phase of cuprates. 

\begin{figure}[ht!]
\centerline{\includegraphics[bb=20 20 500 400,width=0.7\columnwidth]{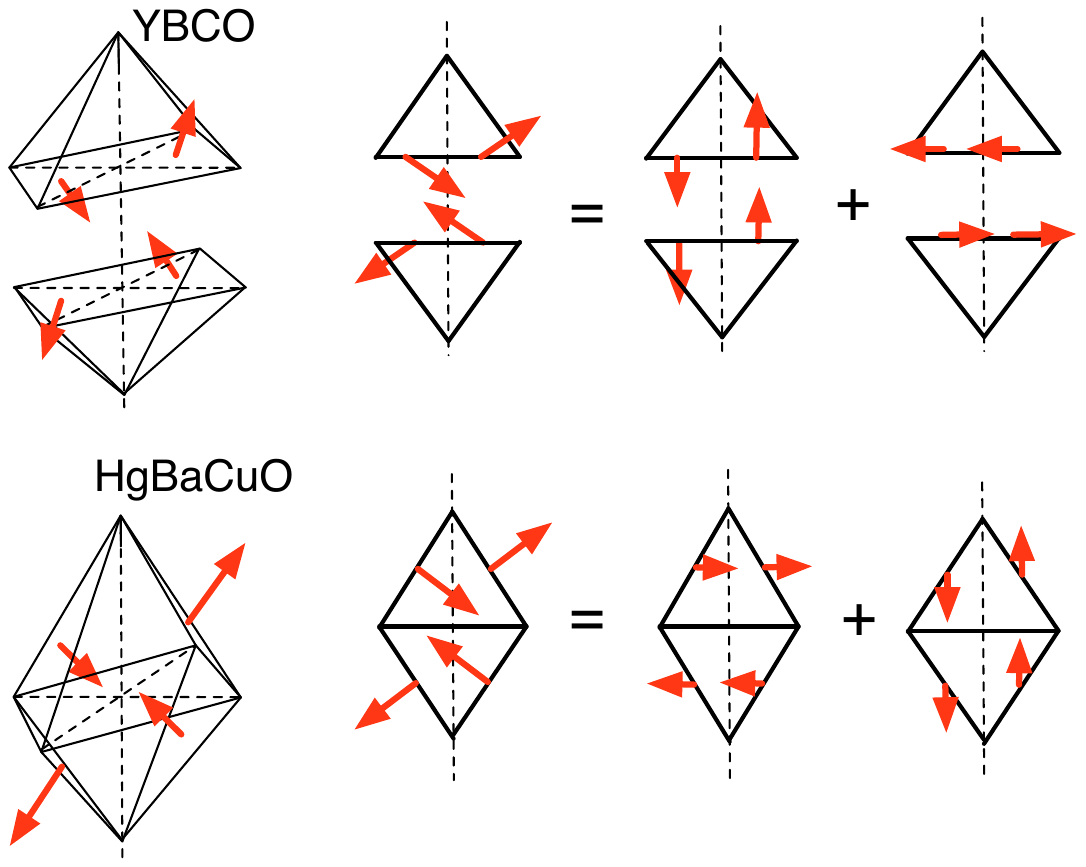}} 
\caption{Schematic representation of the magnetic flux distribution observed in the experiment. The diagram symbolically represents decomposition of the magnetic structure into horizontal and vertical components. The actual composition of the two into the physical order parameter depends on microscopic details and is not dictated by symmetry.}
\label{fig:apical}
\end{figure}

The simplest structure that has non-zero toroidal moment is a pair of moments at finite offset; each moment is directed perpendicular to the line connecting them; the moments are of equal magnitude and opposite in direction. However, function $\bold{M}(\bold{r})$ may have more elaborate structure; see the end of this section for discussion of the experimental situation. 
It is instructive to separate $\bold{M}(\bold{r})$  into into planar and perpendicular components, $\bold{M}(\bold{r})=\bold{M}_{\perp}(\bold{r})+\bold{M}_{\parallel}(\bold{r})$; here  $\bold{M}_{\perp} = \bold{z}M_z$ and $\bold{M}_{\parallel} = \bold{x}M_x+\bold{y}M_y$ and $x,y$ is the basis in the plane parallel to the copper-oxide plane and $z$ is in the direction perpendicular to it.  Fig.~\ref{fig:apical} schematically represents this decomposition with the red arrows understood as a magnetization direction $\bold{M}(\bold{r})$.  Under the point group operations each component, $\bold{M}_{\perp}$ and $\bold{M}_{\parallel}$, transforms independently under its own $E_u$ representation. Consequently, each component may contribute independently to the toroidal moment $\bold{L}=(L_{x},L_{y})$ via Eq.~(\ref{eq:toroidal}). We conclude that on symmetry grounds the magnetic structure in the pseudogap phase can be an arbitrary combination of , $\bold{M}_{\perp}$ and $\bold{M}_{\parallel}$ ; the precise balance between the two depends on the microscopic details and is decided experimentally, see discussion in the end of this section.

Experimental data and theoretical calculations indicate that microscopic orbital currents within the unit cell are responsible for magnetic structure in pseudogap phase of cuprates. Here we discuss symmetry aspects of this relation.   In general, magnetic structure in the crystal can be equivalently described  in terms of the periodic magnetization function,  $\bold{M}(\bold{r})$, or a pattern of periodic microscopic currents $\bold{j}_m$ closed within the unit cell (loop currents). The two are related by  
\begin{align}\label{eq:aaa}
\bold{j}_m= c\curl \bold{M}\,,
\end{align}
see Ref.~\onlinecite{LLP}. In the absence of macroscopic currents, $\bold{H}=\bold{B}+4\pi \bold{M}=0$, the magnetization  $\bold{M}(\bold{r})$  is proportional to the magnetic field within the unit cell, $\bold{B}=-4\pi \bold{M}$. The magnetization can be solved in terms of microscopic currents  $\bold{j}_m(\bold{r})$ by inverting equation  $\bold{j}_m= c\curl\bold{M}$~:
\begin{align}\label{eq:aaa}
\bold{M}(\bold{r}) = \frac1{4\pi{c}}\int d\bold{r}' \frac{\bold{j}_m(\bold{r}')\times(\bold{r}-\bold{r}')}{|\bold{r}-\bold{r}'|^3}
\end{align}
where integral $d\bold{r}'$ is over the whole volume of the crystal. If the magnetization  $\bold{M}(\bold{r})$ is a periodic function on the lattice, so is  $\bold{j}_m(\bold{r})$, and vice-versa. 

We can relate directly the toroidal order parameter $\bold{L}$ to the microscopic current distribution within unit cell. Current pattern is restricted to the unit cell, $\int_{\text{unit cell}} d\bold{r} \bold{j}_m(\bold{r})=0$. Consider second moment of current distribution 
\begin{align}\label{eq:a1}
\int_{\text{unit cell}} d\bold{r} \bold{r}^2  \bold{j}_m(\bold{r})\,.
\end{align} 
Using definition of microscopic current in terms of magnetization, $\bold{j}_m= c\curl\bold{M}$, we write this as 
\begin{align}\label{eq:a2}
\int_{\text{unit cell}} d\bold{r} c \bold{r}^2 \curl\bold{M}(\bold{r})\,. 
\end{align} 
Using identity 
\begin{align}\label{eq:a3}
\bold{r}^2\curl\bold{M} = \curl(\bold{M}\bold{r}^2)+ 2\bold{M}(\bold{r})\times\bold{r}
\end{align} 
and the definition of the toroidal moment in Eq.~(\ref{eq:toroidal}) we find  
\begin{align} 
\bold{L} = \int_{\text{unit cell}} d\bold{r} (1/2c) \bold{r}^2 \bold{j}_m(\bold{r}) - \frac1v\int d\bold{s}\times \bold{M}(\bold{r}) \bold{r}^2/2\,.
\end{align} 
The second integral is over the surface of the unit cell, $d\bold{s}$ is surface element. Using the fact that the surface is shared between adjacent unit-cells, the surface integral vanishes.  Therefore, one can define a toroidal moment by 
\begin{align}\label{eq:aaa}
\bold{L} = \int_{\text{unit cell}} d\bold{r} (1/2c) \bold{r}^2 \bold{j}_m(\bold{r})\,. 
\end{align}
We conclude that the toroidal order parameter is proportional to the planar (parallel to copper-oxide plane) component  of the second moment of microscopic currents within the unit cell.  

The magnitude of the spin-flip intensity observed in experiments for the most underdoped samples studied is consistent with a pair of magnetic moments $\sim0.1\mu_B$ in the centroids of the two triangles; the centroids are located at distance $x_0\sim{a}/2$ from the center of the unit cell ($a$ is unit cell size in the CuO plane).  The magnetic moment  $\bold{M}$ is given by a volume integral of magnetization,  $\int d\bold{r} \bold{M}(\bold{r})$. Integrating over half of unit cell (the full integral over unit cell vanishes) we must obtain magnetic moment of order of $0.1\mu_B$; we conclude that the  magnetization in the unit cell is estimated as $|\bold{M}|\sim 0.1 \mu_B/a^3$ and the integral in Eq.~(\ref{eq:toroidal}) is $\bold{L} \sim |\bold{M}| a^4 \sim 0.1 \mu_B a$. 
\begin{figure}[ht!]
\centerline{\includegraphics[bb=20 20 400 200,width=0.7\columnwidth]{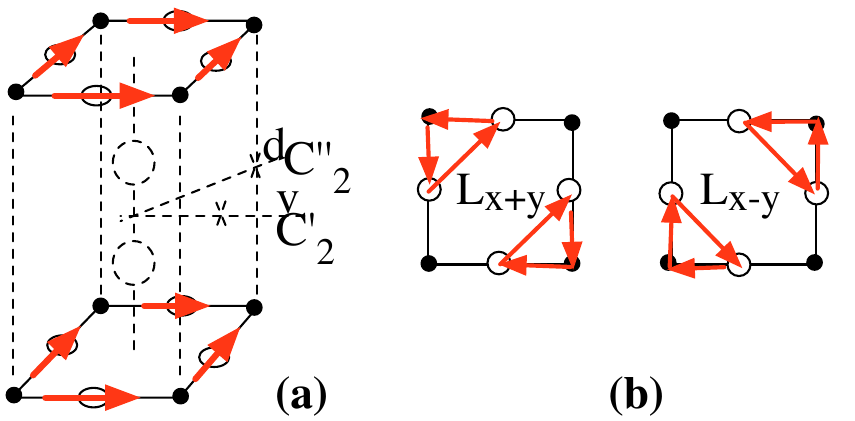}} 
\caption{(a) YBCO unit cell, black dots - copper atoms; open circles - oxygen atoms. Operations of the point group are indicated. (b) Representations of the point group can be constructed on the space of patterns of currents on the inter-atomic links in the unit cell. Two diagrams indicate the pair of current patterns that transforms under inversion-odd two dimensional representation $E_u$. The loop order parameter transforms under the same representation and can be represented by these pictures.}
\label{fig:loop}
\end{figure}

The simplest loop-current pattern that leads to magnetic structure consistent with an in-plane toroidal moment consists of a pair of planar current loops passing via planar oxygens and planar copper, see Fig.~\ref{fig:loop}(b). Such distribution of microscopic currents generates two opposite magnetic fluxes directed along $\bold{z}$ axis, see Fig.~\ref{fig:apical}; as has been mentioned earlier,  this is consistent with the pattern of the magnetization which has non-zero toroidal moment. However, polarized neutron scattering  experiments indicate a finite horizontal component of the magnetization $\bold{M}(\bold{r})$ in the unit cell\cite{NeutronExp,greven}. If the assumption is made in the analysis of the experiments that the structure factor for the in-plane and the out of plane components is the same, the magnitude of the horizontal and the vertical components are similar.  To understand the presence of both horizontal and vertical components in the pattern of magnetization one has to consider also spin-orbit interactions  and/or loop currents involving the apical oxygen.
In YBCO each copper-oxide plane is seeing non-centro-symmetric environment; consequently, symmetry allows a Dzialoshinskii-Moriya type interaction which in the presence of vertical component of the magnetization (planar loop currents) leads to a horizontal component of the magnetization due for example to an in-plane polarization of electron spin in the unit cell \cite{aji-varma}. In mercury compound this mechanism is not allowed because each copper-oxide plane sees centro-symmetric environment. A horizontal component however naturally appears if one considers loop currents via apical oxygen \cite{weber,greven} as well as via planar copper.  It is then reasonable that the major part of the horizontal component in YBCO is also due to loop-currents via apical oxygens.

\section{ Lattice distortions that accompany loop-current order.}
As mentioned earlier \cite{varma2006}, $\underline{m}mm$ symmetry does not allow  piezo-magnetism, i.e. no distortion changing spontaneously the symmetry of the unit-cell to linear order in the order parameter is allowed. Here we discuss lattice deformations that accompany current-loop order to second order. To do so and for the other results derived in this paper, one must specify the irreducible representations of the crystal symmetry in the absence of the loop-current order. The copper-oxide lattice has tetragonal symmetry 4/mmm ($D_{ 4h}$). The operations in the group fall into 10 equivalence classes: $\pi/2$-rotations around $z$  axis $C_4$, $\pi$-rotations around $z$  axis $C_2$, $\pi$-rotations around $x$  axis $C_2^{\prime}$, $\pi$-rotations around $x+y$ axis $C_2^{\prime\prime}$, $\pi/2$-rotations around $z$  axis followed by reflection $S_4$, reflections $\sigma_h,\sigma_v\sigma_d$ in the planes perpendicular to axes $z,x,x+y$ respectively, spatial inversion $i$ and identity operation $E$, see Fig.~\ref{fig:loop}(a). Under symmetry operations in  $D_{4h}$, any physical object transforms under one of the 10 irreducible representations~:
\begin{align} 
\begin{aligned} 
\begin{array}{r||lr|}
A_{1g}&     \;           & 1\;\text{or}\; z^2\;\text{or}\; x^2+y^2   \\
A_{2g}&                  &  xy(x^2-y^2)\\
B_{1g}&                  & x^2-y^2 \\
B_{2g}&                  & xy\\
E_{g} &                  &  (zx, zy)
\end{array}
\end{aligned}
\;, \;\;
\begin{aligned} 
\begin{array}{r||lr|}
A_{1u}&  \;\;                  &   xyz(x^2-y^2) \\
A_{2u}&                        &   z \\
B_{1u}&                        &   xyz \\
B_{2u}&                        &   (x^2-y^2)z  \\
E_{u}  &                       &   (x,y) 
\end{array}
\end{aligned} \;.
\label{eq:facts}
\end{align}
The second column illustrates each irreducible representations by a polynomial of the same symmetry. Within $D_{4h}$ the vector representation of the rotations in space is no more irreducible; instead, the the polar vector $\boldsymbol{E}$ and axial vector $\boldsymbol{M}$ break into irreducible representations of $D_{4h}$ as follows~:
\begin{align}
\begin{array}{r|| l}
A_{2g}& \qquad M_z                       \\
E_{g} & \qquad (M_{y'}, M_{x'})     \\
A_{2u}& \qquad E_z                             \\
E_{u}  & \qquad (E_{y'},E_{x'})             
\end{array}
\label{eq:vectors}
\end{align}
The basis $x,y$ is chosen along the copper-oxide links. We use notation $x',y'$ for a basis rotated by $45^{\circ}$ with respect to $x,y$, i.e., $x'=(x-y)/\sqrt2$ and $y'=(x+y)/\sqrt2$. Representations $E_u$ and $E_g$ are  two-dimensional representations, all other irreducible representations are one-dimensional. 

\begin{figure}[ht!]
\centerline{\includegraphics[bb=20 20 450 470,width=0.8\columnwidth]{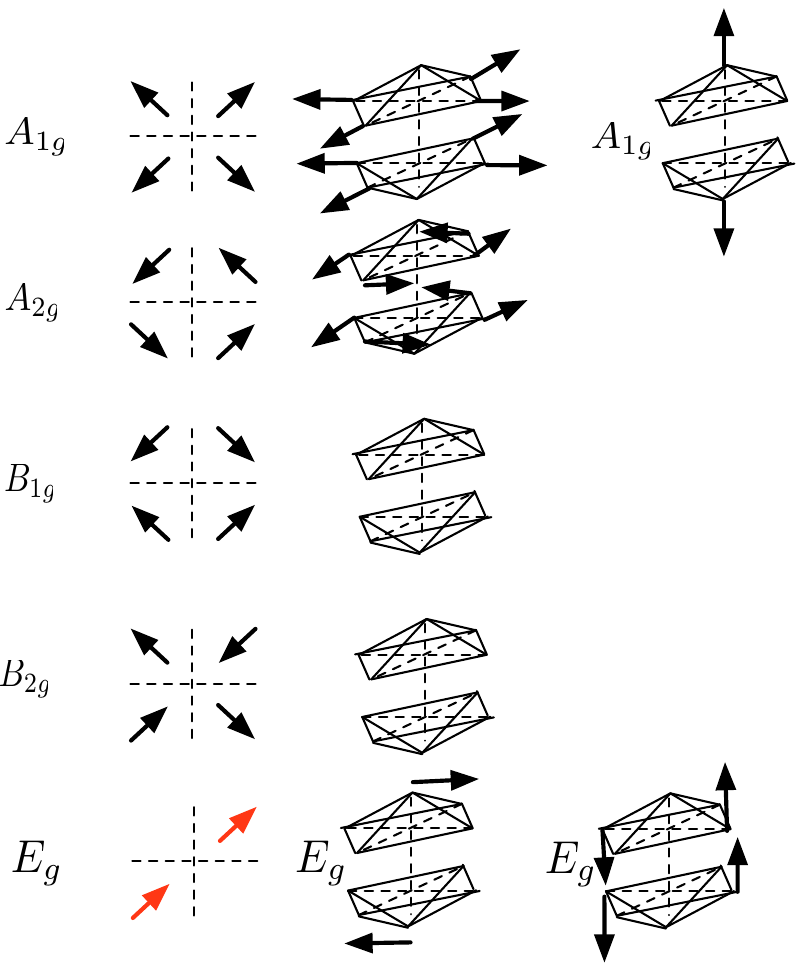}} 
\caption{  Inversion-even representations of the group $D_{4h}$. Black arrows represent a polar vector, such as atom displacement; red arrow represent an axial vector, such as magnetization.}
\label{fig:irred-even}
\end{figure}

The term in the free energy that can couple loop order and lattice distortion has to be of at least a second order in the current-loop order parameter since $\bold{L}$ is time-reversal-odd; such a term has a structure $L L u$ where $u$ describes distortion. As a consequence, the lattice distortion is of a second order in loop-order parameter, $u\propto L^2$. To obtain the symmetry of allowed distortions we have to break bilinear products of $\bold{L}$ into irreducible representations. Within $D_{4h}$ the product of two polar in-plane vectors (transforming under  $E_u$ representation) breaks into four irreducible representations~:
\begin{align}
\begin{array}{r||cl}
A_{1g}  &\qquad         &{L}_{y'}^2+{L}_{x'}^2  \\
A_{2g}  &\qquad        & {L}_{y'}{L'}_{x'}-{L}_{x'}{L'}_{y'}\\
B_{1g}    &\qquad        & {L}_{y'}{L}_{x'}\\
B_{2g}     &\qquad        & {L}_{y'}^2-{L}_{x'}^2
\end{array} 
\label{eq:even} 
\end{align}
Let us consider each of the terms above. One kind of distortions that belongs to $A_{1g}$ variety amounts to a change in the size of the cell and is obviously allowed to couple to the square of the order parameter. 
Examples of the distortions in  the other irreducible representations are shown in Fig~\ref{fig:irred-even} where black arrows represent atom displacements. The most interesting case is that corresponding to the third and fourth lines above: the free energy contains a term $\gamma_{B_{2g}} u_{B_{2g}} ( {L}_{y'}^2-{L}_{x'}^2)$ where $u_{B_{2g}}$ is the distortion that belongs to ${B_{2g}}$ irreducible representation and $\gamma_{B_{2g}}$ is a coupling constant; similarly for $B_{1g}$.  From the point of view of symmetry, this distortion is reminiscent of how finite polarization is generated in {\it improper ferroelectrics} \cite{LLP}.

In YBCO distortion of the $A_{2g}$ symmetry is allowed in the presence of loop order and intrinsic orthorhombic distortion. At low enough temperatures the crystal structure of YBCO has intrinsic orthorhombic distortion  $u(B_{1g})$ of the  $B_{1g}$ symmetry due to the ordering in the copper-oxigen chains (the lattice constants are $a = 3.82$\AA, $b =3.89$\AA).  Within  $D_{4h}$ group the following relation holds $B_{1g}B_{2g} = A_{2g}$ (this can be deduced from the character table). Therefore, there exists a  term $u(A_{2g}) u(B_{1g})  [{L}_{y'}^2-{L}_{x'}^2] $ which will create a distortion $u(A_{2g})$ of $A_{2g}$ symmetry in the presence of finite orthorombic lattice distortion $u(B_{1g})$ and loop order parameter $L_{x'}$ or $L_{y'}$. Another effect of finite orthorhombic  distortion $u(B_{1g})$ in YBCO is to introduce a mixing between $L_{y'}$ and $L_{x'}$ via the term $ u(B_{1g}) L_{y'}L_{x'}$. Along the same line of analysis one concludes that no distortion of $A_{2g}$ symmetry is allowed in mercury compound in the pseudogap phase since its lattice has tetragonal symmetry. 

The magnitude of the distortions may be estimated from the  free energy 
\begin{align}
\delta F = \gamma (u/a) L^2 + E (u/a)^2/2
\end{align} 
where $u$ is distortion in a particular symmetry channel ($a$ is in-plane lattice constant), $L^2$ stand for one of the loop-order bilinears in Eq.~(\ref{eq:even})), and $E$ is the elastic modulus (in energy units) for the distortion in the corresponding symmetry channel. Minimizing the Free energy we obtain $u/a = \gamma L^2/E$. 
It is quite difficult to get an estimate of the magnitude of the $\gamma$'s directly. For an order of magnitude estimate, we proceed as follows: The $A_{1g}$ distortion arises from the same symmetry arguments as the change in the volume of ferromagnetic substances due to their order and we may use experimental results obtained for them to get an idea of the order of magnitude of the effect to be expected. For example, detailed measurements are available for $Fe$ \cite{stuart-ridley}. The lattice constant changes from above the transition at over 800 degrees C to 300 degrees C due to ferromagnetism with a moment of about 2.5 $\mu_B$ per atom by less than a part in $10^3$. With the ordered moment an order of magnitude smaller, we would therefore expect for similar coupling constant and Bulk modulus, a change in similar temperature region of only a part in $10^5$. This estimate is unlikely to be incorrect by more than an order of magnitude. One would expect that similar magnitude of distortion is to be expected in the interesting case of the $B_{2g}$ distortion and much smaller for the $A_{2g}$ distortion since that must rely also on the orthorhombicity of the original structure.

\section{Deformations induced by magnetic field.}

\begin{figure}[ht!]
\centerline{\includegraphics[bb=20 20 450 450,width=0.8\columnwidth]{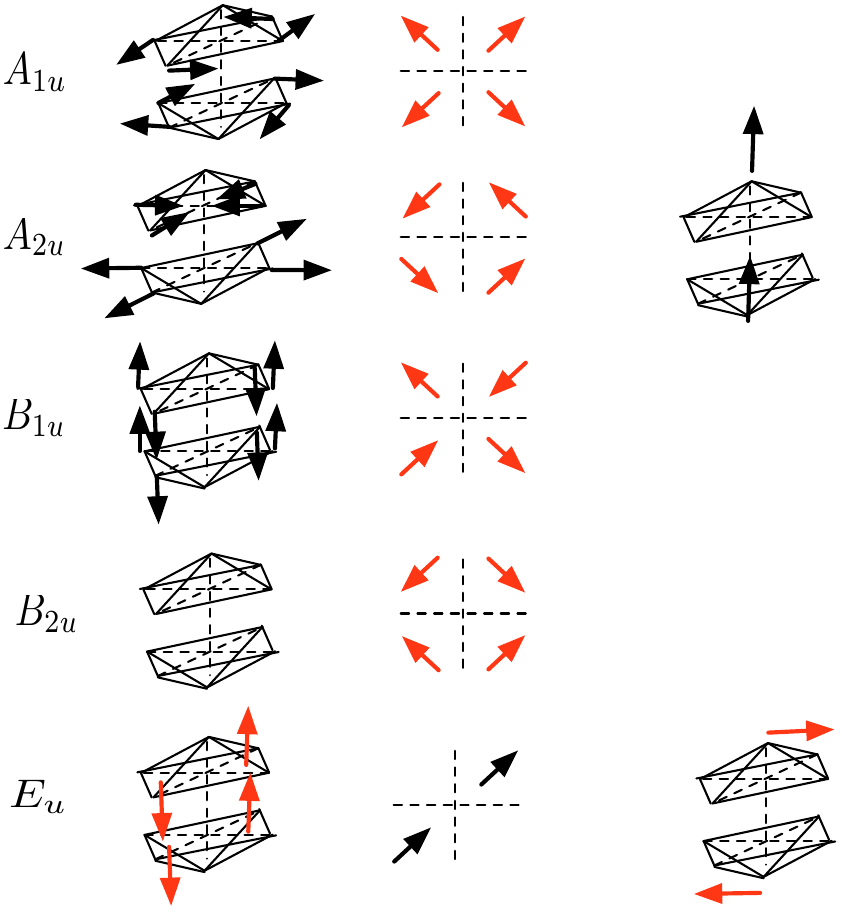}} 
\caption{ Inversion-odd representations of the group $D_{4h}$.} 
\label{fig:irred-odd} 
\end{figure}

We now consider possible unit cell distortions that can couple linearly to magnetic field $H$ and linearly to the order parameter. Such distortion is allowed by symmetry, since the order parameter $L$ is polar time-reversal-odd vector. So a term in the free-energy $\propto uLH$ is allowed with the lattice distortion $u$ being odd under space inversion. To find the symmetry of all allowed distortions we must decompose the product of $\boldsymbol{L}$ and $\boldsymbol{H}$ into irreducible representations. For vector $\boldsymbol{H}=(H_{x'},H_{y'})$ in-plane (transforming under $E_g$), the product $L H$ breaks into four irreducible representations
\begin{align}
\begin{array}{r||ll}
A_{1u}&  \qquad                         & {L}_{y'}{H}_{y'}+{L}_{x'}{H}_{x'}   \\
A_{2u}&  \qquad                         & {L}_{y'}{H}_{x'}- {L}_{x'}{H}_{y'}  \\
B_{1u}&  \qquad                         & {L}_{y'}{H}_{y'}- {L}_{x'}{H}_{x'}    \\
B_{2u}&  \qquad                         & {L}_{y'}{H}_{x'}+{L}_{x'}{H}_{y'}  
\end{array} 
\label{eq:odd-Hparallel}
\end{align}
For magnetic field out of plane $\boldsymbol{H}=H_{z}$ (which transforms under $A_{2g}$)  the distortion  $\boldsymbol{u} = (u_{x'},u_{y'})$  transforms under $E_u$. Indeed, the product $u L$ contains irreducible representation  $A_{2g}$~: 
\begin{align}
\begin{array}{r||ll}
A_{2g}&  \qquad                         & {L}_{y'}{u}_{x'}- {L}_{x'}{u}_{y'}  \\
\end{array} 
\label{eq:odd-Hperp}
\end{align}
We conclude that depending on the orientation of the magnetic field and the order parameter, distortion can appear in each of the five inversion-odd representations, see Fig~\ref{fig:irred-odd}.

To estimate the magnitude of the distortion we consider free energy  
\begin{align}\notag
\delta F = \beta (\frac{u}a) L M + E (\frac{u}a)^2/2
\end{align} 
where $u$ is distortion in one of the possible irreducible representations and $L M$ is the corresponding bilinear. Minimization gives  $u/a = \beta L M/E$. We can estimate the magnitude of the coefficient $\beta$ as well as typical distortion magnitude using data from recent magnetization measurements \cite{Monod2008}. The cubic term of the form 
$\beta (\frac{u}a) L M$ introduces additional contribution to magnetic susceptibility via the term $C L^2 M^2$ which is obtained upon minimization with respect to $u/a$; here $C=\beta^2/E$. The change in susceptibility in the presence of loop order is $\delta \chi \sim - \chi^2 C L^2$, see Ref~\onlinecite{Gronsleth} for details. Both $L$ and $M$ can be measured in units of Bohr magnetons per unit cell, with $L$ interpreted as staggered magnetization within unit cell; the convenient energy units are set with $\mu_B^2/\AA^3 \sim 1K$. In the experiment the susceptibility is measured $~3 mm^3/Mol$ (or about $3\times10^{-5}$ in dimensionless units for YBCO) at temperatures about pseudogap temperature. Since the loop order saturates rather quickly below the pseudogap temperature, we can attribute experimental value $\delta \chi\sim 0.1\chi$ to saturated loop order magnitude $L_s\sim 0.1 \mu_B/\text{Unit cell}$. We estimate $C \sim (\delta\chi/\chi^2)(1/L_s^2) \sim 10^8 K^{-1}/\text{Unit cell}$; the unit cell volume of YBCO is $\sim170\AA^3$. Using $C=\beta^2/E$ with $E\sim 10 eV/\text{Unit cell}$ we estimate $\beta\sim 5\times10^6$ in dimensionless units (where $L^2$ and $M^2$ are in units of energy per volume of unit-cell). We can now estimate the magnitude of the distortion. In the experiment the magnetization at external field of $1T$ corresponds to magnetic moment $M\sim 10^{-5}\mu_B/\text{Unit cell}$.  In this situation the distortion is estimated as  $u/a = \beta L M/E \sim 10^{-6}$.  

\subsection{Magnetic field dependence of elastic neutron scattering intensity.}

We now discuss a possibility of detecting  lattice distortions in the present of external magnetic field  by the usual x-ray or neutron crystallographic techniques.  The dominant contribution to elastic neutron scattering intensity comes from contact interaction between neutron and atomic nucleus. One can analyze these starting with   
\begin{align}\label{eq:amplit}
& I_N(q) \propto |A_N(q)|^2 \notag\\
&A_N(q) =  \sum_{R,\alpha}  f_{\alpha} e^{i\bold{q} (R+\bold{r}_{\alpha})}
\end{align}
where $q$ is (neutron) momentum transfer. $\bold{r}_{\alpha}$ is position of atom $\alpha$ within unit cell $R$;  $f_{\alpha}$ are nuclear scattering amplitudes. Let us assume that due to lattice distortion atoms shift to a new positions $\bold{r}_{\alpha} \rightarrow \bold{r}_{\alpha}+\bold{u}_{\alpha}$ within unit cell and ask whether there is a change in intensity to linear order in $\bold{u}_{\alpha}$~:
\begin{align}\label{eq:change}
\delta I_N(q) = \delta A_N^{\dagger}(q) A_N(q) + c.c 
\end{align}
For centro-symmetric crystal the sum in Eq.~(\ref{eq:amplit}) evaluates to a real number (the nuclear amplitudes $f_{\alpha}$ are real for $q$ typical of slow neutrons), i.e., the amplitude  $A_N(q)$ is real. To linear order in $\bold{u}_{\alpha}$ the change in the amplitude is
\begin{align}\label{eq:amplit-change}
\delta A_N(q) = \sum_{R,\alpha}  f_{\alpha} i\bold{q} \bold{u}_{\alpha} e^{i\bold{q} (R+ \bold{r}_{\alpha})}
\end{align}
In loop-ordered phase, external magnetic field can produce several types of distortions, all of which are odd under spatial inversion, see Eqs.~(\ref{eq:odd-Hparallel}) and (\ref{eq:odd-Hperp}). For the inversion-odd distortion  $\bold{u}_{\alpha}$ the sum in the right-hand side of Eq.~(\ref{eq:amplit-change}) is pure imaginary; intensity variation $\delta I_N(q)$  given by Eq.~(\ref{eq:change}) vanishes. We conclude that the lattice distortion associated with external magnetic field, though non-zero, does not change intensity of the Bragg peaks in elastic neutron scattering to linear order in magnetic field.

\section{Magneto-electric effects in the pseudogap phase}

Copper-oxide metal in the pseudogap phase exhibits magneto-electric effects. Magnetic field out of plane belongs to $A_{2g}$ irreducible representation. The product of an in-plane electric field $\bold{E}$ and loop order $\bold{L}$ has a component $\bold{E}\times\bold{L}$ that belongs to the same,  $A_{2g}$  irreducible representation (and similarly for out-of plane electric field which belongs to  $A_{2u}$)~:
\begin{align}
\begin{array}{r||l|l}
A_{2g}& \;\; H_z                & \;\; \bold{E}\times\bold{L} = {E}_{y'}{L}_{x'}-{E}_{x'}{L}_{y'}\\
A_{2u}& \;\; E_z                & \;\; \bold{H}\times\bold{L}= {H}_{y'}{L}_{x'}- {H}_{x'}{L}_{y'}  \\
\end{array} 
\end{align}
Corresponding to the two lines in this table, there are two terms in the free energy that couple electric and magnetic fields in the presence of loop order parameter $\bold{L}= (L_{x'},L_{y'})$~:
\begin{align}\label{eq:magnetoelectricterm}
\delta F_{\text{ME}} = &\lambda_1\, H_z\,(E_{y'} L_{x'} - E_{x'} L_{y'})  \notag\\
&\qquad +\lambda_2\,  E_z\, ( H_{y'} L_{x'} - H_{x'} L_{y'} ) \,.
\end{align}
We rewrite these in a matrix form~:
\begin{align}\label{eq:magnetoelectric}
\left[\begin{array}{ccc}  H_{x'} & H_{y'} & H_z  \end{array}\right]
\left[\begin{array}{ccc} 
0 & 0 &  -\lambda_2 L_{y'}\\
0  & 0 &  \lambda_2 L_{x'} \\
-\lambda_1 L_{y'} & \lambda_1 L_{x'} & 0 
\end{array}\right]
\left[\begin{array}{c}  E_{x'} \\ E_{y'} \\ E_z  \end{array}\right] \,.
\end{align}
The two coupling constants $\lambda_{1,2}$ have different magnitude since they describe coupling in two different irreducible  representations. Assuming a single domain phase, say $\bold{L} = (L_{x'},0)$, we obtain a magneto-electric tensor in the symmetry-broken phase of the form~:
\begin{align}\label{eq:magnetoelectrictensor}
\left[\begin{array}{ccc} 
0 & 0 &  0\\
0  & 0 &  \alpha_2  \\
0 & \alpha_1  & 0 
\end{array}\right]\,.
\end{align}
where $\alpha_{1,2} = \lambda_{1,2}L_x$. General symmetry consideration allow axial time-reversal-odd tensor of the second rank of this form in the crystal class $\underline{m}mm$, see Ref.~\onlinecite{Birss}. 

We now discuss the magnitude of the magneto-electric susceptibility $\alpha$ in cuprates that is associated with the magnetic structure in the pseudogap phase. In spin systems the magneto-electric coefficient is limited by the magnitude of the spin-orbit coupling \cite{Rado1962}; a typical value $\alpha\sim10^{-4}$ to $10^{-3}$ (dimensionless in cgs units) is several orders of magnitude smaller than the upper bound set by thermodynamics \cite{Brown1968}. In the cuprates the magnetic structure in the pseudogap phase has an orbital origin. However due to finite resistivity of cuprates in the pseudogap phase the observation of the magneto-electric phenomena may not be possible. To estimate the magnitude of the magneto-electric coefficient $\alpha$ we start with thermodynamic identity (see, e.g., Ref.~\onlinecite{Rivera1994})
\begin{align}\label{eq:aaa}
\alpha_{ij} = -\frac{\partial^2F}{\partial E_i\partial H_j} = \frac{\partial M_j}{\partial E_i}  = \frac{\partial P_i}{\partial H_j}
\end{align}
We use the last equality, i.e., we apply magnetic field $H$ and estimate the induced polarization. It has been estimated earlier that in the experiments in Ref.~\onlinecite{Monod2008} the displacement of copper and oxygen atoms relative to each other is of order $u/a\sim10^{-6}$ at external field of  $1T$; the symmetry of the displacement has been discussed earlier in the paper. The polarization which corresponds to such displacement is estimated as $P\sim (e/a^2) (u/a)\sim 10^{-3} esu$ (here $a$ is the ab plane unit cell size $\sim4\AA$; we have assumed that copper and oxygen in the copper-oxide plane have charge $\pm e$). We find that magneto-electric susceptibility is $\alpha=P/H\sim 10^{-5}$, (which is dimensionless in cgs units). This corresponds to about $3 \times 10^{-8}$ Gauss/(Volt/cm). This value could be observed experimentally in an insulator but not in the metallic pseudogap state of the cuprates where final voltage arises in the bulk only due to finite resistivity. We should mention for completeness that the orbital ordered phase also supports a zero-field Hall effect. But for a field applied of $10$ volts/cm, the Hall effect would correspond only to that due to a magnetic field of about $10^{-6}$ Gauss.

Not all effects accompanying the observed orbital order are so small. For example, x-ray dichroism is predicted to be observable \cite{dimatteo}. So is the predicted second harmonic generation \cite{simon}. All the effects investigated here come due to the periodic magnetic fields generated by the orbital-ordered phase, which are always small. The effective electronic energy of such phases is two to three orders of magnitudes larger than the magnetic energies. This is similar to the difference between exchange energies and magnetic field energies in magnetic order due to spin-moments.

\acknowledgements{Numerous discussions with Vivek Aji, Philippe Bourges and Steve Kivelson are greatly acknowledged. The suggestion of a distortion in the ordered phase proportional to an applied magnetic fields was made to us by P.W. Anderson. The suggestion of the monoclinic distortion due to the orbital order worked out here is originally due to Steve Kivelson. A.S. would like to acknowledge hospitality of the Aspen Center of Physics where part of the work was done.}

\end{document}